\documentclass[cameraready]{Interspeech}

\title{UNet-Based Fusion and Exponential Moving Average Adaptation for Noise-Robust Speaker Recognition}

\author[affiliation={1}]{Chong-Xin}{Gan}
\author[affiliation={2}]{Peter}{Bell}
\author[affiliation={1}]{Man-Wai}{Mak}
\author[affiliation={3}]{Zhe}{Li}
\author[affiliation={1}]{Zezhong}{Jin}
\author[affiliation={1}]{Zilong}{Huang}
\author[affiliation={1}]{Kong Aik}{Lee}

\address{
    $^1$ The Hong Kong Polytechnic University, Hong Kong SAR, China \\
    $^2$ Centre for Speech Technology Research,  The University of Edinburgh, United Kingdom \\
    $^3$ The University of Hong Kong, Hong Kong SAR, China
}

\email{chong-xin.gan@connect.polyu.hk}

\keywords{Speech enhancement, speaker recognition, exponential moving average, UNet feature fusion}
\usepackage{comment}
\usepackage{cite}
\usepackage{tabularx}
\usepackage{amsmath}
\usepackage{bm}
\usepackage{booktabs}
\usepackage{xcolor, multirow, makecell}
\usepackage[table]{xcolor}

\usepackage{comment}

\begin{document}

\maketitle

\begin{abstract}
    The joint training of speech enhancement and speaker embedding networks for speaker recognition is widely adopted under noisy acoustic environments. While effective, this paradigm often fails to leverage the generalization and robustness benefits inherent in large-scale speech enhancement pre-training. Moreover, maintaining the speaker information in the denoised speech is not an explicit objective of the speech enhancement process. To address these limitations, we proposed a scalable \textbf{U}Net-based \textbf{F}usion framework (UF-EMA) that considers the noisy and enhanced speech as a multi-channel input, thereby enabling the speaker encoder to exploit speaker information effectively. In addition, an \textbf{E}xponential \textbf{M}oving \textbf{A}verage strategy is applied to a speaker encoder pre-trained on clean speech to mitigate overfitting and facilitate a smooth transition from clean to noisy conditions. Experimental results on multiple noise-contaminated test sets showcase the superiority of the proposed approach. 
\end{abstract}

\section{Introduction}

The advent of deep neural networks (DNNs) has recently transformed speaker verification (SV) \cite{ mak2020machine, li2026towards}. In contrast to the traditional i-vector approaches \cite{dehak2010front}, DNN-based methods have shown outstanding speaker modeling capabilities, thereby facilitating the extraction of discriminative speaker features for robust speaker recognition \cite{li2025mutual, desplanques20_interspeech, snyder2018x}. The availability of large-scale annotated speech datasets \cite{Chung18b, ko2017study, snyder2015musan, Nagrani17} further enhances the performance and generalization of SV systems by diversifying speaker characteristics and simulating complicated acoustic environments. When optimized with advanced loss functions \cite{li2022real, jung2022pushing}, these systems achieve state-of-the-art performance.

Despite these advances, the sole training of a speaker encoder for SV tasks is still vulnerable to background interference and distortion\textemdash such as noise, channel effects, and overlapping speakers\textemdash in real-world conditions. Severe background noise hinder the speaker encoder from capturing speaker-dependent cues, and overlapping speakers cause the speaker encoder to extract features from the wrong speaker. To mitigate this, a common approach is to cascade a speech enhancement (SE) model with the speaker encoder, i.e., performing speech denoising prior to extracting the speaker representation \cite{sun2023noise, ma2024gradient, kim22b_interspeech, gao22c_interspeech}. This cascade approach allows the speaker embedding network to map the enhanced speech to robust speaker representations, thereby reducing the negative impact caused by noise. Based on the training strategies, current methods can be broadly divided into two categories: separate training \cite{ plchot2016audio, eskimez2018front} and joint training \cite{gao22c_interspeech, sun2023noise, kim22b_interspeech, shi20_odyssey, dowerah2023joint}.

Separate training involves either using a pre-trained SE model for denoising and training a speaker encoder independently, or training the SE model and speaker encoder separately. Specifically, the noisy speech is first mapped to its clean counterpart using a conventional speech enhancer pre-trained for noise suppression, followed by passing the resulting enhanced signal to a speaker encoder for embedding extraction \cite{plchot2016audio, eskimez2018front}.

Conversely, in joint training, a network is optimized to simultaneously achieve the objectives of SV and SE. For example, a mean squared error (MSE) loss is employed in \cite{kim22b_interspeech} to ensure that the enhanced speech generated by the SE module can closely match the clean speech, and simultaneously, a classification loss minimizes the discrepancy between the predicted and true speaker identities. Another representative method, VoiceID loss \cite{shon19b_interspeech}, guides the SE model to generate a ratio mask that filters out irrelevant components from noisy spectrograms. Because the regression loss tends to saturate earlier than the classification loss, the authors in \cite{gao22c_interspeech} propose stopping the optimization of the SE model once overfitting in SE is detected, but updating the SV model continuously. Furthermore, it has been reported that speaker-dependent information can be better preserved by enabling the enhanced spectrogram to approximate the gradient of the clean spectrogram \cite{ma2024gradient, li2021gradient}.
\begin{figure*}[ht] 
    \centering
    \includegraphics[scale=0.61]{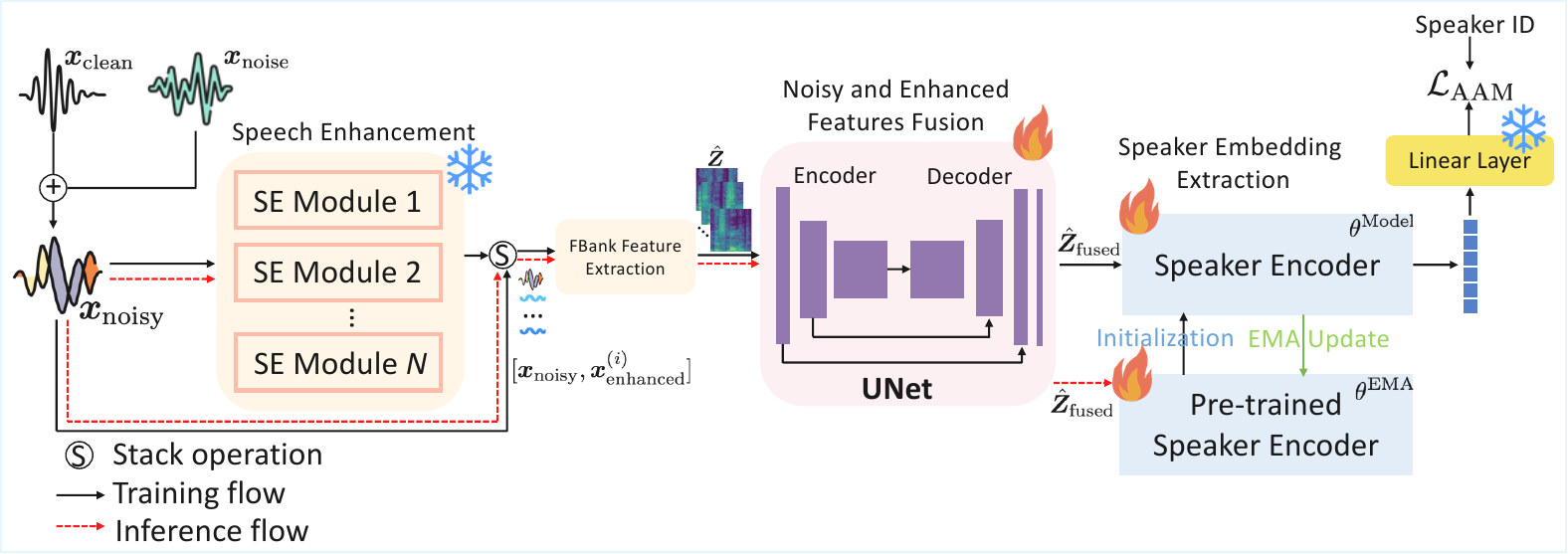}
    \caption{Overview of the proposed UF-EMA method. After data augmentation, the noisy speech $x_\mathrm{noisy}$ is inputted to $N$ pre-trained SE models. The spectrograms of the resulting enhanced speech signals and the original noisy waveform are fed into a UNet-based fusion module, generating a fused spectrogram $z_\mathrm{fused}$ for the speaker encoder. The speaker encoder is initialized with a pre-trained SV model and updated in an exponential moving average (EMA) manner. The entire system is optimized with the AAM loss using the ground truth speaker labels. We used two pre-trained SE models ($N=2$) in our experiments.} 
    \label{fig:fig1}
\end{figure*}
While these methods yield promising results, they produce enhanced speech solely from noisy waveforms, overlooking speaker-specific information embedded within the noisy input. Moreover, these methods train the SE models from scratch, missing the opportunity to leverage pre-trained SE networks that have undergone extensive speech denoising training.

The primary challenge in employing a pre-trained SE module lies in the potential introduction of undesirable artifacts. In automatic speech recognition (ASR), this issue is often mitigated by applying observation addition (OA) \cite{sato2022learning} as a post-processing step that integrates the noisy waveform into the recognition process. For instance, the authors in \cite{10715447} linearly combined noisy and enhanced speech based on their signal-to-noise ratio (SNR) to improve ASR performance. To capture richer information beyond the signal level, a ResNet-based bridge module was proposed to jointly process noisy and enhanced waveforms \cite{cui2025reducing}. However, both approaches assume that the fused speech can be represented as a linear interpolation of noisy and enhanced signals, which may lose valuable information. Motivated by this limitation, we propose employing a UNet architecture to effectively capture and integrate complementary information from both sources, thereby generating an optimally fused spectrogram.   

Prior works on SE-SV either train the speaker embedding network from scratch or keep it frozen. However, directly feeding enhanced speech into the speaker encoder may result in distorted embedding vectors, as the discrepancies between the clean and enhanced speech may degrade the model's speaker representation capability. Conversely, freezing the encoder can compromise its robustness to noise. To tackle this trade-off, we introduced an exponential moving average (EMA) strategy to update the pretrained speaker extractor, ensuring a smooth and gradual transition from clean to noisy environments, which can further improve the overall performance.  

In summary, this paper makes two primary contributions. First, to alleviate the adverse effects introduced by the pretrained SE model and to exploit the speaker-dependent information within noisy speech, a \textbf{U}Net is employed to \textbf{F}use the noisy and enhanced signals at the spectrogram level, tailoring a well-suited input for speaker verification. Second, to prevent catastrophic forgetting and enhance the noise robustness of the speaker encoder, we adopt an \textbf{E}xponential \textbf{M}oving \textbf{A}verage strategy to dynamically update the parameters in the speaker encoder. The proposed UF-EMA delivers outstanding results, outperforming many existing approaches.

\section{Methodology}

An overview of the proposed framework is illustrated in Fig.~\ref{fig:fig1}. The noisy speech is first generated by mixing clean utterances with various types of noises through data augmentation. Subsequently, several pre-trained speech enhancement (SE) models are employed in parallel to denoise the speech and produce enhanced speech signals. To mitigate potential artifacts introduced by the SE models and to fully leverage both the noisy and enhanced speech, a UNet-based fusion module is utilized to generate a more robust representation of the input speech. Finally, a dynamic optimization strategy is applied to the speaker encoder, enabling it to gradually enhance its robustness to noise.
\subsection{Speech Denoising}
Given the noisy speech signal $\bm{x}_\mathrm{noisy}$ with length $L$, which is a mixture of the clean speech signal $\bm{x}_\mathrm{clean}$ and background noise $\bm{x}_\mathrm{noise}$, the objective of a speech enhancement network is to estimate an enhanced version $\bm{x}_\mathrm{enhanced}$ that exhibits improved speech intelligibility and perceptual quality. 
The network is typically optimized using a loss function that measures the discrepancy between the enhanced and clean signals, such as the mean squared error (MSE).

In recent years, a wide range of deep learning-based approaches to SE have been developed, leading to significant improvement in speech denoising performance. The SE models are typically pre-trained on large-scale datasets such as the Deep Noise Suppression (DNS) corpus and utilize multi-level signal representations encompassing spectral magnitude, phase, and temporal features. Recent methods \cite{deoliveira22_interspeech, chao2024investigation, luo2023music} adopt diverse network designs to capture rich contextual information, thereby achieving more effective noise suppression while better preserving speech fidelity.

To leverage the strengths of these pre-trained models, we incorporate them into our framework for speech denoising. Because different denoising approaches tend to look at different aspects of the speech signal, incorporating multiple SE modules can be beneficial. In this work, we adopt two powerful SE models: band-split recurrent neural network (BSRNN) \cite{luo2023music} and DEMUCS \cite{defossez20_interspeech}. BSRNN was pre-trained in the spectral domain, where a spectrogram was partitioned into several non-overlapping subbands that were processed independently by dedicated blocks in the network. The resulting features are then merged to construct a time-frequency mask used to suppress noise and interference. In contrast, DEMUCS is an encoder–decoder architecture trained in an end-to-end manner to operate directly on raw waveforms.

We employ both SE modules simultaneously, allowing their enhanced outputs to provide complementary cues that are beneficial for downstream speaker representation learning. It is worth noting that the number of SE models can be scaled up to $N$, forming an ensemble of SE networks to further improve the robustness and generalization of the denoising process. Formally, the enhancement procedure is expressed as:
\begin{equation}
    \bm{x}_\mathrm{enhanced}^{(i)} = f^{(i)}(\bm{x}_\mathrm{noisy}), \quad i \in \{1, 2, \ldots, N\}
\end{equation}
where $\bm{x}_\mathrm{enhanced}^{(i)}$ represents the $i$-th signal outputted by the $i$-th SE module.

\subsection{Noisy and Enhanced Speech Fusion}

Prior studies commonly feed only the enhanced speech into the speaker encoder, disregarding the original noisy signal. This practice overlooks the substantial speaker-discriminative information contained in the noisy speech. Moreover, it is well recognized that artifacts introduced by speech enhancement models can distort speaker-related information, thereby degrading recognition performance. To alleviate this issue, many studies \cite{sato2022learning, 10715447} employ observation addition (OA) as a post-processing step, wherein a dynamic coefficient is estimated to linearly interpolate the noisy and enhanced waveforms. However, such a linear combination fails to capture the complex nonlinear relationships and complementary characteristics inherent in the noisy and enhanced speech signals. To address this limitation, we employ a UNet-based fusion network that transforms the noisy-enhanced pairs into a robust fused representation, which is subsequently fed into a speaker encoder.

As shown in the middle part of Fig.~\ref{fig:fig1}, after the denoising process, the enhanced speech $\bm{x}_\mathrm{enhanced}^{(i)}$ from each SE module $f^{(i)}$ is stacked with the noisy speech $\bm{x}_\mathrm{noisy}$ along the channel dimension, forming a multi-channel input $\hat{\bm{x}}  \in \mathbb{R}^{(N+1) \times L}$:
\begin{equation}
\hat{\bm{x}} = \operatorname{stack}(\bm{x}_\mathrm{noisy}, \bm{x}_\mathrm{enhanced}^{(1)}, \bm{x}_\mathrm{enhanced}^{(2)}, \ldots, \bm{x}_\mathrm{enhanced}^{(N)}),
\end{equation}
where $\operatorname{stack}(\cdot)$ represents the stack operation along the channel dimension.

Before passing $\hat{\bm{x}}$ to the fusion network, log-mel filter bank features are extracted, resulting in a feature representation $\hat{\bm{Z}}  \in \mathbb{R}^{(N+1) \times T \times F}$. Here, $T$ and $F$ denote the number of frames and frequency bins, respectively. The fusion module is implemented by following a standard encoder-decoder design. Identical to the frame-level extractor in ResNet34, the encoder within the UNet contains several convolutional blocks. Each block consists of two 2D convolution layers and a batch-normalization layer, followed by a ReLU activation function. 

The decoder comprises several deconvolution blocks, in which the feature maps from the encoder are progressively upsampled to restore the original time-frequency resolution. To alleviate overfitting, skip connections are incorporated between the encoder and decoder layers. The final output of the decoder is a fused representation that combines complementary information from both the noisy and enhanced speech, effectively mitigating distortions caused by speech enhancement. This process can be expressed as:
\begin{equation}
\hat{\bm{Z}}_\mathrm{fused} = g_\theta(\hat{\bm{Z}}),
\end{equation}
where $g_\theta(.)$ represents the UNet-based fusion network and $\hat{\bm{Z}}_\mathrm{fused} \in \mathbb{R}^{T \times F}$ denotes the UNet's output.

Overall, the proposed method enables the UNet to act as a fusion module combining the noisy and enhanced speech, producing a unified and robust representation that can be utilized by the downstream speaker encoder.
\subsection{Exponential Moving Average of Speaker Encoder}

Conventional approaches often freeze the parameters of the speaker encoder after pre-training. Although this strategy can stabilize the training process, it heavily relies on the front-end SE module to generate optimal acoustic representations, which is not always available in practice. In contrast, the joint training of the speaker encoder and the SE module often complicates optimization, as the two components exhibit different convergence behaviors.

To overcome these challenges, we initialize the speaker encoder with the pre-trained weights of a speaker encoder and update its parameters using an exponential moving average (EMA) strategy. The update rule is defined as:
\begin{equation}
\theta_{t+1}^\mathrm{EMA} = \alpha \theta_{t}^\mathrm{EMA} + (1 - \alpha){\theta}_t^\mathrm{Model},
\label{eq.4}
\end{equation}
where $\theta_t^\mathrm{EMA}$ denotes the smoothed parameters at iteration $t$, and ${\theta}_t^\mathrm{Model}$ represents the current model parameters obtained via standard gradient updates. At $t = 0$, $\theta_{0}^\mathrm{EMA}$ and $\theta_{0}^\mathrm{Model}$ are initialized with the pre-trained weights of the speaker encoder. Then, $\theta_1^{\mathrm{Model}}$ is updated via classification loss, and Eq.~\ref{eq.4} is applied to update $\theta^{\mathrm{EMA}}_{1}$. $\alpha \in [0, 1)$ is the smoothing coefficient that controls the update rate. This mechanism allows the speaker encoder to gradually adapt from clean to noisy environments while retaining the discriminative speaker information acquired during pre-training.
\section{Experimental Settings}
The development set of VoxCeleb1 \cite{Nagrani17} was utilized as the training data, while Vox1-O was employed for evaluation. We randomly truncate speech files into 2-second segments. When SE was not applied, the 80-dimensional log-mel filter banks were extracted from the speech features and used as input to the speaker encoder. When SE was applied, the fused spectrograms from the output of the UNet were applied to the speaker encoder.
\begin{table*}[ht]
\caption{Comparisons with existing works in terms of EER (\%). The baseline in Row 1 corresponds to the evaluation results of the pretrained speaker encoder. The row under Noise, Music, and Babble specifies the SNRs (in dB) of the test speech. The best and second-best results are highlighted in bold face and underline, respectively.}
\centering
\small
\setlength{\tabcolsep}{4pt}
\begin{tabularx}{0.95\textwidth}{c | c | c | *{3}{>{\centering\arraybackslash}X} | *{3}{>{\centering\arraybackslash}X} | *{3}{>{\centering\arraybackslash}X} | >{\centering\arraybackslash}X}
\toprule
\multirow{2}*[-0.5ex]{Row} & \multirow{2}*[-0.5ex]{Method} & \multirow{2}*[-0.5ex]{Clean} & \multicolumn{3}{c}{Noise} & \multicolumn{3}{c}{Music} & \multicolumn{3}{c|}{Babble} & \multirow{2}*[-0.5ex]{Average}\\
\cmidrule(lr){4-6} \cmidrule(lr){7-9} \cmidrule(lr){10-12}
& & & 0 & 5 & 10 & 0 & 5 & 10 & 0 & 5 & 10 & \\
\midrule
1 & Baseline & 3.00 & 8.56 & 5.73 & 4.67 & 8.90 & 5.49 & 4.20 & 13.80 & 6.01 & 4.32 & 6.47\\
2 & NDML \cite{sun2023noise} & 2.90 & 10.24 & 6.96 & 5.02 & 10.84 & 6.52 & 4.66 & 10.96 & 6.13 & 4.28 & 6.85\\
3 & Cai \textit{et al.} \cite{cai2020within} & 3.12 & 7.34 & 5.65 & 4.35 & 7.79 & 5.23 & 4.11 & 11.78 & 5.97 & 4.44 & 5.98\\
4 & VoiceID \cite{shon19b_interspeech} & 2.61 & 6.38 & 4.64 & 3.87 & 6.35 & 4.38 & 4.36 & 9.45 & 4.76 & 3.59 & 4.94 \\
5 & NDAL \cite{xing24_interspeech} & 2.63 & \underline{5.87} & \underline{4.19} & 3.53 & 7.07 & 4.77 & 3.70 & \textbf{6.80} & \underline{4.43} & 3.66 & 4.67 \\
6 & ExU-Net \cite{kim22b_interspeech} & 2.76 & 6.80 & 5.23 & 4.07 & 7.35 & 4.90 & 3.69 & 9.57 & 5.52 & 4.06 & 5.40 \\
7 & Diff-SV \cite{kim2024diff}  & \textbf{2.35} & 6.01 & 4.52 & \underline{3.49} & \underline{6.04} & \underline{3.96} & \textbf{3.10} & 8.74 & 4.51 & \textbf{3.33} & \underline{4.61} \\
8 & Cho \textit{et al.} \cite{cho2025multi} & - & 6.50 & 4.71 & 3.73 & 6.46 & 4.37 & 3.48 & 8.22 & 4.64 & 3.53 & 5.07 \\
9 & UF-EMA (proposed) & \underline{2.55} & \textbf{5.36} & \textbf{4.01} & \textbf{3.35} & \textbf{5.04} & \textbf{3.90} & \underline{3.24} & \underline{7.01} & \textbf{4.36} & \underline{3.40} & \textbf{4.22} \\
\bottomrule
\end{tabularx}
\label{table:1}
\end{table*}
To simulate diverse acoustic environments, we followed the setup in \cite{kim22b_interspeech} where the MUSAN dataset \cite{snyder2015musan} was randomly divided into two non-overlapping subsets: one used for augmenting the training data and the other for contaminating the test data, thereby preventing data leakage. The learning rate was set to 1e-3, and the momentum parameter $\alpha$ was set to 0.999. The test set was further corrupted with noise, music, and babble at signal-to-noise ratios (SNRs) of $-5$, 0, 5, and 10 dB. During training, room impulse responses (RIRs) \cite{ko2017study} and the training subset of MUSAN were applied for data augmentation.

For speech enhancement, BSRNN \cite{yu23b_interspeech} and DEMUCS \cite{defossez20_interspeech} were employed due to their proven effectiveness in speech separation \cite{li2024effectiveness} and enhancement tasks \cite{wang2021tstnn,cao2022cmgan}. During both training and inference, their parameters remained frozen. For the fusion network, the encoder adopts the same frame-level architecture as ResNet34, taking 80-dimensional mel-filter bank features as input and consisting of four convolutional blocks with output channel sizes of 32, 64, 128, and 256, respectively. The decoder progressively upsampled the feature maps to the original resolution.

The ECAPA-TDNN \cite{desplanques20_interspeech} was employed as the speaker encoder, and the AAM Softmax was used as the loss criterion. The dimension of speaker embedding was set to 192. EER was used to evaluate the performance.

\section{Results and Discussions}

\subsection{Main Results}
Table~\ref{table:1} presents a comprehensive comparison of the proposed method with the existing speaker verification approaches under clean and noisy conditions, including noise, music, and babble, at SNRs of 0, 5, and 10 dB. Under the clean condition, the proposed method delivers a competitive EER of 2.55\%. Although Diff-SV \cite{kim2024diff} achieves a lower EER of 2.35\%, it relies on a diffusion model that iteratively reconstructs clean features from noisy inputs, leading to a substantially higher computational cost during inference. In contrast, our method achieves comparable performance with a simpler and more efficient design.
\begin{table}[ht]
    \centering
    \caption{Ablation study on different components of the proposed UF-EMA method under $-5$ dB SNR.}
    \resizebox{0.91\columnwidth}{!}{
    \begin{tabular}{c l c c c}
        \toprule
        Row & Components & Noise & Music & Babble \\
        \midrule
        1 & All & \textbf{7.66} & \textbf{9.50} & 17.04 \\
        2 & w/o Noisy input & 8.49 & 10.89 & 22.57 \\
        3 & w/o BSRNN & 8.49 & 9.68 & 18.68 \\
        4 & w/o DEMUCS & 8.48 & 9.96 & \textbf{15.74} \\
        5 & w/o EMA (Fixed) & 9.13 & 11.41 & 22.38 \\
        6 & w/o EMA (From scratch) & 7.97 & 9.92 & 18.05 \\
        7 & w/o EMA (Fine-tune) & 7.78 & 9.75 & 17.43 \\
        
        \bottomrule
    \end{tabular}}
    \label{tab:my_label2}
\end{table}

For a fair comparison, the UNet component in VoiceID \cite{shon19b_interspeech} was replaced by the UNet employed in our framework. Comparing Row 4 and Row 9, the proposed method consistently outperforms VoiceID across all conditions. Compared with ExU-Net, our approach demonstrates superior performance. Overall, the proposed method achieves either the lowest or the second-lowest EER across all evaluation settings and ranks first in terms of average EER.
\subsection{Effect of Individual Components}
Table~\ref{tab:my_label2} presents an ablation study evaluating the contribution of different components of the proposed method under the challenging conditions in which the clean test utterances were corrupted by noise, music, and babble interference at an SNR of $-5$ dB. Row 2 shows that removing the noisy input results in a substantial performance drop, with EERs rising from 7.66\%, 9.50\%, and 17.04\% to 8.49\%, 10.89\%, and 22.57\%, respectively. This indicates that rich speaker-dependent information in the original noisy speech can contribute to speaker discrimination.

Rows 3 and 4 show that omitting the enhanced features outputted by BSRNN or DEMUCS degrades performance, suggesting the effectiveness of incorporating the enhanced features extracted by these pre-trained SE modules. Table 2 also shows that among the three types of noise, babble consistently yields the poorest performance. We conjecture that in these low-SNR babble scenarios, the pretrained SE model may mistakenly attenuate the target speaker’s speech, as it struggles to discriminate the target from interfering speakers. Results also show that the EMA strategy plays a pivotal role because the EER was increased significantly without this strategy. 
\begin{figure}
    \centering
\includegraphics[width=0.95\linewidth]{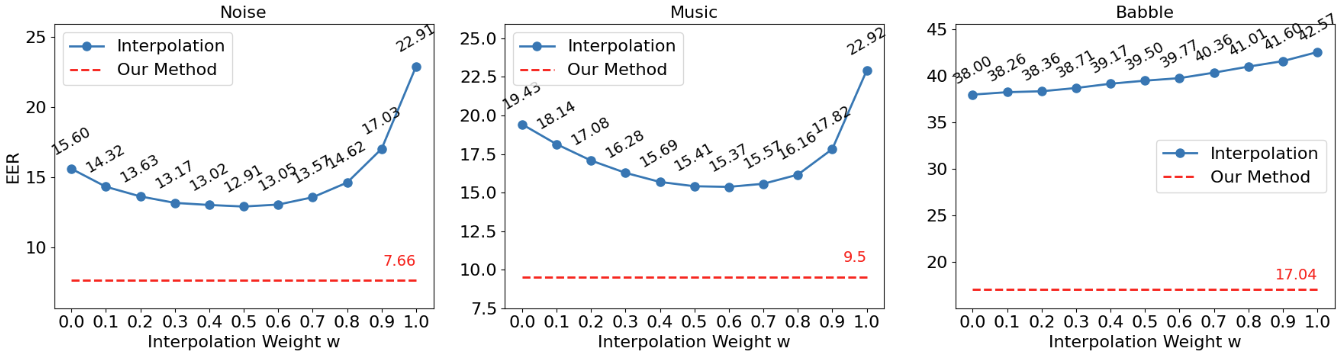}
    \caption{Comparing the proposed method with linear interpolation of noisy and enhanced speech under noise, music, and babble at $-5$ dB SNR.}
    \label{fig:placeholder}
\end{figure}
\subsection{Comparison with Linear Interpolation of Noisy and Enhanced Speech}
Instead of using the UNet to fuse the noisy and enhanced features, we may also interpolate them using an interpolation weight $w \in [0,1]$ such that $x_{\mathrm{fused}}=wx_{\mathrm{enhanced}}+(1-w)x_{\mathrm{noisy}}$. Figure~\ref{fig:placeholder} shows how the EERs vary when the contribution of noisy and enhanced speech varies. It can be observed that the performance gradually improves when the weight increases from 0.0 to 0.5. Beyond this point, the EER increases sharply. A similar phenomenon can also be observed under music contamination. This behavior suggests that, under low SNR, the enhanced speech and the original noisy speech can complement each other to improve speaker verification performance. However, linear interpolation can only achieve limited performance improvement, with a high risk of getting poorer performance than not having speech enhancement. By contrast, the proposed method combines enhanced and noisy speech in a more effective manner, demonstrating superior performance. 
\section{Discussion}
We here proposed a robust speaker verification framework that integrates pretrained speech enhancement models. Using a UNet-based fusion network, the system effectively combined noisy and enhanced speech to improve robustness. To ensure a smooth adaptation from clean to noisy conditions, EMA was applied to the speaker encoder, further stabilizing the training. Experimental results across a wide range of acoustic conditions demonstrated the effectiveness and superiority of the proposed method in handling challenging noisy environments while maintaining competitive performance on clean speech.

\bibliographystyle{IEEEtran}
\bibliography{mybib}

\end{document}